\documentclass[%
reprint,superscriptaddress,amsmath,amssymb, floatfix,aps,prc,twocolumn]{revtex4-2}

\usepackage{graphicx}
\usepackage{subcaption}
\usepackage{color}
\usepackage{dcolumn}
\usepackage{bm}

\usepackage[normalem]{ulem}  
\usepackage{orcidlink}
\usepackage{hyperref}
\hypersetup{breaklinks=true, colorlinks=true, citecolor=blue}

\newcommand{\MeV}{\text{MeV}}
\newcommand{\G}{\text{G}}

\begin{document}


\title{Effect of Magnetic Fields on Urca Rates in Neutron Star Mergers}

\author{Pranjal Tambe \orcidlink{0000-0003-2293-6953}}
 \email{pranjal.tambe@iucaa.in}
 \affiliation{%
 Inter University Centre for Astronomy and Astrophysics, Ganeshkind, Pune 411007, India
}%
\author{Debarati Chatterjee \orcidlink{0000-0002-0995-2329}}%
 \email{debarati@iucaa.in}
\affiliation{%
 Inter University Centre for Astronomy and Astrophysics, Ganeshkind, Pune 411007, India
}%


\author{Mark Alford \orcidlink{0000-0001-9675-7005}}
\affiliation{%
 Physics Department, Washington University, St. Louis, MO 63130, USA
}%
\author{Alexander Haber \orcidlink{0000-0002-5511-9565}}
 \email{ahaber@physics.wustl.edu}
\affiliation{%
 Physics Department, Washington University, St. Louis, MO 63130, USA
}%

\date{\today}

\begin{abstract}
Isospin-equilibrating weak processes, called ``Urca" processes, are of fundamental importance in astrophysical environments like (proto-)neutron stars, neutron star mergers, and supernovae. In these environments, matter can reach high temperatures of tens of MeVs and be subject to large magnetic fields. We thus investigate Urca rates at different temperatures and field strengths by performing the full temperature and magnetic-field dependent rate integrals for different equations of state. We find that the magnetic fields play an important role at temperatures of a few MeV, especially close to or below the direct Urca threshold, which is softened by the magnetic field. At higher temperatures, the effect of the magnetic fields can be overshadowed by the thermal effects.
We observe that the magnetic field more strongly influences the neutron decay rates than the electron capture rates, leading to a shift in the flavor equilibrium. 

\end{abstract}

\maketitle
\section{Introduction}
\label{sec:intro}

Our knowledge of the behaviour of dense matter is limited to terrestrial laboratory physics, such as in nuclear laboratories and heavy-ion experiments in particle accelerators. Neutron Stars (NS),  compact astrophysical objects containing the densest form of matter in the universe, may however allow us to probe matter in extreme environments, at densities beyond nuclear saturation density ($n_0$) in the highly isospin asymmetric regime. The rich pool of available multi-messenger astrophysical data makes NSs unique laboratories to constrain theories of nuclear physics for ultradense, cold and highly neutron-rich matter~\cite{Shapiro:1983du, Glendenning, Schaffner-Bielich_2020}.

For comparison with astrophysical data, the global structure of NSs can be obtained by solving the Tolman-Oppenheimer-Volkoff (TOV) equations, i.e.~a set of coupled general relativistic hydrostatic equations, supplemented with a relation between pressure and energy density for the neutron star matter, the equation of state (EoS)~\cite{Burgio:2021vgk, Lattimer:2021emm}. The EoS depends on the nature of interactions between matter constituents, and can be modeled in various ways.
One approach is to perform ab-initio calculations of realistic nucleon-nucleon interactions treating many-body problems using Brueckner-Hartree-Fock (BHF) or Dirac-BHF calculations~\cite{Baldo:1991zz, VanGiai:2010tm, Sammarruca:2019ncy, Burgio:2021vgk}. Another approach uses phenomenological models 
with an effective interaction whose coupling constants are fit to the properties of finite nuclei, the saturation properties of infinite nuclear matter, hypernuclear experimental data, or additionally to neutron matter calculations from chiral effective field theory \cite{Alford:2022bpp, Alford:2023rgp, Salinas:2023nci, Drischler:2021kxf, Drischler:2024ebw}.

In recent years, there have been many observations of NS masses and radii that impose constraints on the EoS parameters, such as from pulse profile modeling of pulsars as well as from gravitational wave (GW) observations and their electromagnetic (EM) counterparts \cite{Graham2022, LIGOScientific:2017vwq, LIGOScientific:2018cki, LIGOScientific:2018hze, LIGOScientific:2020aai, LIGOScientific:2021qlt, Pang:2021jta, Ascenzi:2024wws}. These studies provide improved estimates of the highest possible mass of a neutron star, $M_{max}>2.072_{-0.066}^{+0.067}$ M$_{\odot}$ \cite{NANOGrav:2019jur}, or on the radius of a $1.4$ M$_{\odot}$ neutron star $R_{1.4}=11.98_{-0.40}^{+0.35}$ km \cite{Riley:2019yda, Miller:2019cac, Riley:2021pdl, Miller:2021qha, Salmi:2022cgy, Pang:2022rzc, Choudhury:2024xbk, Vinciguerra:2023qxq}, which have important implications on dense matter theories.

 Most neutron stars are observed to be cold compared to the Fermi energies of their constituent particles, allowing us to often employ the zero temperature approximation. 
 However, there are certain scenarios involving neutron stars where the temperature can be much higher and may play a non-negligible role in the dynamics. After their formation in core-collapse supernova, newly born neutron stars can have temperatures as high as $50$ MeV but quickly cool 
below temperatures of $1$ MeV \cite{Pons:1998mm, Lentz:2015nxa, Burrows:2020qrp, Janka:2006fh}. Recent binary neutron star (BNS) merger simulations have shown that matter in mergers can reach temperatures as high as $80$ MeV~\cite{Hanauske:2019qgs, Most:2022wgo, Perego:2019adq, Raithel:2022nab}, the matter near the core of neutron stars reaching temperatures up to several tens of MeV. In such scenarios, thermal effects on the EoS cannot be ignored \cite{Raithel:2019gws,Raithel:2021hye}.

Urca processes play an important role in flavor equilibration and controlling neutron star dynamics, such as cooling, or producing bulk viscosity that could result in damping of density oscillations in binary neutron star mergers or unstable $r$-modes that are potential sources of continuous GW~\cite{Lattimer:1991ib,Haensel:2000vz, Yakovlev:2000jp, Shternin:2018dcn, Schmitt:2017efp, Haskell:2015iia}. In cold neutron stars, NS matter is assumed to be in flavor equilibrium. 
However, in scenarios such as newly born hot neutron stars (i.e.~proto-neutron stars or PNS) or the binary neutron star (BNS) merger remnant, matter can be driven out of flavor equilibrium via density oscillations on the millisecond timescale \cite{Baiotti:2016qnr,Alford:2017rxf}, and flavor equilibrium is reestablished by Urca processes. 
This relaxation process
can give rise to bulk viscous dissipation of density oscillations. If the timescale of viscous damping processes is comparable to the timescale of BNS mergers then it can have a significant impact on the merger dynamics \cite{Alford:2017rxf, Alford:2019qtm, Most:2021zvc, Most:2022yhe, Alford:2024tyj}.

At temperatures up to about $5\,\MeV$ nuclear matter is neutrino-transparent: neutrinos escape freely because their mean free path is larger than the radius of the star. In this regime the processes
that equilibrate the proton fraction are neutron decay
$n\to p_e^-+\bar\nu_e$ (nd) and electron capture $p+e^- \to n+\nu_e$ (ec).
Flavor equilibrium is defined by 
equality of the forward and backward reaction rates, $\Gamma_\text{nd}=\Gamma_\text{ec}$.
At low temperatures (less than $1\, \MeV$) and magnetic fields the energy carried off by neutrinos is negligible, and in this approximation the equilibrium condition becomes
\begin{equation}
\mu_n=\mu_p+\mu_e ~.
\label{eq:cold-beta-eq}
\end{equation}
However the forward and backward processes are not the time reverse of each other, so this is not an exact equilibrium condition. At the upper end of the neutrino-transparent regime ($1\,\MeV\lesssim T\lesssim 5\,\MeV$) the neutrino energy becomes non-negligible, resulting in significant corrections to
\eqref{eq:cold-beta-eq}~\cite{Alford:2018lhf,Alford:2021ogv}.

This is significant because this temperature range is where dissipative  effects arising from flavor equilibration processes are most important.
The dependence of the bulk viscous dissipation time on density and temperature as well as its sensitivity to the EoS has been investigated systematically in several works for $npe$ matter, as well as including muons and hyperons, for both matter in neutrino-transparent and neutrino-trapped regimes~\cite{Alford:2019kdw,Alford:2020lla,Alford:2021lpp,Alford:2020pld,Alford:2022ufz, Alford:2023uih}. It was demonstrated that at $T \sim 4-6$ MeV the resonant maximum of the bulk viscosity occurs, while in the neutrino-trapped regime, i.e., temperatures well above $5\,\MeV$, the bulk viscosity decreases $\zeta \sim T^{-2}$ and drops sharply by several orders of magnitude compared to the value in the neutrino-transparent regime.

In this work we study how magnetic fields affect flavor equilibration (specifically, direct Urca) rates.
Neutron stars typically have strong magnetic fields $B \sim 10^9-10^{12}\,\G$. In a subclass of neutron stars called magnetars, ultra-strong magnetic fields as large as $10^{15}-10^{16}\,\G$ has been observed at the surface, and could potentially be higher in their interior \cite{Mereghetti:2015asa, Kaspi:2017fwg, Esposito:2018gvp, Konar:2017kty}. Neutron star simulations indicate that the remnants produced as a result of binary neutron star mergers can be strongly magnetized \cite{Harding:2006qn, Lorimer:2008se, Kiuchi:2015sga, Ciolfi:2017uak, Ciolfi:2020cpf}. The presence of ultra-strong magnetic fields can influence the EoS and structure of neutron stars, see, e.g., Ref.~\cite{Chatterjee:2021wsr} and the references
therein. Urca processes in presence of magnetic fields have been previously studied in~\cite{Lai1991, Leinson:1998yr, Baiko:1998jq, Maruyama:2021ghf, Anand:2000ud, Sinha:2008wb}. These studies neglect the finite temperature effects on the Urca process rates and use the Fermi-surface approximation wherein particles only on the Fermi surface participate in the reactions.

In this work, we aim to incorporate the finite temperature effects in calculating the rates of direct Urca processes in presence of magnetic fields by relaxing the Fermi-surface approximation and integrating over the whole phase-space of interacting particles.
We investigate possible  magnetic field corrections to the flavor equilibrium condition by calculating the mismatch between $\Gamma_\text{nd}$ and $\Gamma_\text{ec}$
as a function of baryon density, temperature $T$ (in the range $1{-}5\,\MeV$), and magnetic field $B$.  We
always fix the proton fraction using \eqref{eq:cold-beta-eq}, and check
how big the magnetic field would have to be for it to make a significant contribution to the mismatch of rates.
If its contribution is comparable to the thermal one then it plays an important role in determining what proton fraction corresponds to flavor equilibrium. 

We will perform calculations in 4 regimes of temperature and magnetic field, three of which provide confirmation that our formalism agrees with previously published results. The fourth case, where temperature and magnetic field effects are both non-trivial, is the main result of this paper.

The structure of this paper is as follows. The formalism for the calculation of direct Urca rates at finite temperature and magnetic field is discussed in Sec.~\ref{sec:formalism}.  In Sec.~\ref{sec:checks} we discuss the limiting cases where our formalism reproduces known results. In Sec.~\ref{sec:results} we present our results for the regime of MeV-range temperature and non-negligible magnetic field. Finally in Sec.~\ref{sec:discussions}, we discuss the implications of our findings.

We use natural units where $\hbar=c=k_B=1$ throughout.

\section{Formalism}
\label{sec:formalism}

For this study we use two finite-temperature relativistic mean field models, TMA \cite{Toki:1995ya} and QMC-RMF3 \cite{Alford:2022bpp, Alford:2023rgp}, from the CompOSE repository \cite{Typel:2013rza}. 
For simplicity, we only consider nucleons and electrons, but muons or other baryons can be added in a straightforward manner.
Both EoSs satisfy state-of-the-art NS observational constraints. The maximum mass predicted by TMA EoS is $2.02 M_{\odot}$ and by QMC-RMF3 EoS is $2.15 M_{\odot}$ consistent with observational constraints $M_{max}>2.072_{-0.066}^{+0.067} M_{\odot}$ for PSR J0740+6620 obtained in~\cite{Riley:2021pdl}. The radius for $1.4 M_{\odot}$ star for QMC-RMF3 is $R=12.21$ km and that for TMA is $12.1$ km consistent with $R_{1.4}=11.98_{-0.40}^{+0.35}$ km obtained in \cite{Pang:2022rzc}. The TMA EoS has a direct Urca threshold at  $n_B=2.1 n_0$, while the QMC-RMF3 EoS has no threshold in the range of densities considered in our study. We choose these two EoSs to specifically observe the magnetic field effects on the direct Urca rates above and below the direct Urca threshold densities. 

In Sec.~\ref{sec:rates} below, we outline the formalism to calculate the general formula for direct Urca rates at finite temperature and magnetic fields. In Sec.~\ref{sec:low_field_rates}, we then derive the simplified expression for the rates for the case when the magnetic fields are low corresponding to a large number of filled Landau levels.

\subsection{Rates in the presence of finite temperature and magnetic field}
\label{sec:rates}
To calculate the direct Urca rates at finite temperature in presence of non-zero magnetic fields, we apply the formalism of Baiko and Yakovlev (1998)~\cite{Baiko:1998jq}, assuming the magnetic field along the z-axis with the vector potential as $\vec{A} = B y \hat{x}$. The neutron decay rate in the presence of magnetic field is given by:
\begin{align}
        \Gamma_\text{nd} &= \frac{eB}{(2\pi)^7} \sum_{l l' s_p s_n} \int d^3\mathbf{k}_{n} d^3\mathbf{k}_{\nu} dk_{pz} dk_{ez}\, \lvert M \rvert^2\,\notag \\ &f_n (1-f_p) (1-f_e) \delta (E_n - E_p - E_e -E_{\nu}) \label{eq:rate-nd-general}\\& \delta (k_{nz} -k_{pz} -k_{ez} -k_{\nu z})\ , \notag       
\end{align} 
 where, the particle distribution functions are given by the  Fermi-Dirac distributions  $f_i = 
\left[1 + \exp((E_i-\mu_i)/T)\right]^{-1}$ and the matrix element is
\begin{align}
  \frac{\lvert M \rvert^2}{G^2} &= 2g_A^2(\delta_{s_p,1}\delta_{s_n,-1}F^2_{l',l}(u) + \delta_{s_p,-1}\delta_{s_n,1}F^2_{l',l-1}(u)) \notag\\
  &+ \frac{1}{2} \delta_{s_p,s_n}\left[(1+g_A s_p)^2 F^2_{l',l}(u)\right]\\& + \frac{1}{2} \delta_{s_p,s_n}\left[(1-g_A s_p)^2 F^2_{l',l-1}(u)\right] \,,\nonumber\\[1ex]
  u&\equiv \dfrac{q^2}{2eB}, \quad\text{where}\quad\vec{q}=\vec{k}_{n\perp}-\vec{k}_{\nu \perp} \ .  \label{eq:u-def}
\end{align}
Here $G^2 = G^2_F$ cos$^2(\theta_c)$, $G_F$ is the Fermi coupling constant and $\theta_c$ is the Cabibbo angle, with $G^2=1.1\times 10^{-22}$ MeV$^{-4}$, $g_A=1.26$ is the axial vector coupling constant and
 $F_{l',l}$ is the normalized Laguerre function \cite{Kaminker1981},
\begin{equation}
    F_{l',l}(u) = \sqrt{\frac{l'!}{l!}} u^{(l-l')/2} e^{{-u}/{2}} L_{l'}^{l-l'}(u)\, = (-1)^{l'-l} F_{l,l'}(u) .
    \label{eq:lag_f}
\end{equation}
The indices $l, l'$ label the electron and proton Landau level numbers respectively, and $L_{l'}^{l-l'}(u)$ are Laguerre polynomials. 
We ignore the contributions of particle magnetic moments to their energies.
The dispersion relations for the particles are :
\begin{align}
    E_e &= \sqrt{m^2_e + k^2_{ez} + 2\,eB\,l}\, , \label{eq:Ee}\\
    E_p &= \sqrt{m^{*2}_p + k^2_{pz} + 2\,eB\,l'} + U_p \, , \label{eq:Ep}\\
    E_n &=\sqrt{m^{*2}_n + k^2_{n}} + U_n\, .\label{eq:En}\
\end{align}
where $m^*_N$ is the effective nucleon mass and $U_N$ is the vector self-energy for the nucleon \cite{Hempel:2014ssa}.The sum $\sum_{s_p,s_n}\lvert M\rvert^2$ over spins can be performed if we neglect the contribution of magnetic moments of particles to their energies,
\begin{equation}
   \sum_{s_n,s_p}\lvert M \rvert^2 \approx G^2 (1+3g_A^2)(F^2_{l',l}(u)+F'^2_{l',l-1}(u)) \, .
   \label{eq:matrix}
\end{equation}
In the neutrino transparent regime, neutrinos only carry momentum of order $T$, which is small compared to the Fermi momenta of all other particles. Thus, we neglect the neutrino momentum in the delta-function and the definition of $\vec{q}$. This leads to a simpler form for the rate and the auxiliary variable $u$,
\begin{align}
    \Gamma_\text{nd}&=\frac{G^2(1+3g_A^2)eB}{32\pi^6} \int \,d^3\mathbf{k}_{n}\, dk_{pz}\, dk_{ez}\, \Theta(E_{\nu})E_{\nu}^2\, \notag \\ &\delta (k_{nz} -k_{pz} -k_{ez})\,\sum_{l,l'} (F^2_{l',l}(u)+F'^2_{l',l-1}(u))\,\\& f_n (1-f_p)(1-f_e) \, \notag, \\
    u&= \frac{k^2_{nx}+k^2_{ny}}{2eB} = \frac{k_{n\perp}^2}{2eB} \ .
\end{align}
The neutrino energy $E_{\nu}=E_n-E_p-E_e$ is fixed by energy conservation. 
Integrating over the neutron momentum orientation gives
\begin{align}
    \Gamma_\text{nd}=&\frac{G^2(1+3g_A^2)eB}{16\pi^5} \int \,dk_n\, dk_{pz}\, dk_{ez}\, E_{\nu}^2\, k_n\notag\Theta(E_{\nu})\\& \Theta(k_n-\lvert k_{pz}+k_{ez}\rvert)\,\sum_{l,l'} (F^2_{l',l}(u)+F'^2_{l',l-1}(u)) \notag\\& f_n (1-f_p)(1-f_e) \,\label{eq:Rates_full}.    
\end{align}

We will evaluate this rate in four regimes of magnetic field and temperature. 
\begin{enumerate}
    \item $T\ll (E_{F_p}-U_p - m^*_p), eB\ll (E_{F_p}-U_p)\,T$. The Fermi-surface approximation is valid (i.e. momentum integrals  can be treated as sharply peaked near the Fermi surface) and magnetic field effects are negligible.
    \item $T\ll (E_{F_p}-U_p - m^*_p)$ and $eB \gtrsim (E_{F_p}-U_p)\, T$:
   Fermi surface approximation is valid, magnetic field effects are non-negligible.
   \item $T\gtrsim (E_{F_p}-U_p - m^*_p)$ and $eB\ll (E_{F_p}-U_p)\,T$: 
   Fermi-surface approximation is not valid (momentum integrals must be performed without any approximations), magnetic field effects are negligible.
   \item $T\gtrsim (E_{F_p}-U_p - m^*_p)$ and $eB\gtrsim (E_{F_p}-U_p)\,T$: 
   Fermi-surface approximation is not valid, magnetic field effects are non-negligible.
   \end{enumerate}

The first three of these have been previously studied in the literature, so our calculations will act as consistency checks on our formalism.
The fourth case, where magnetic field effects and departures from the Fermi surface approximation are both significant, is the main result of this paper.

\subsection{Low magnetic fields}
\label{sec:low_field_rates}

When calculating rates in the presence of low magnetic fields wherein
the discretization of the energy into Landau levels is small compared to the Fermi momenta and a large number of Landau levels are occupied, the double sum over large number of Landau levels in Eq.~(\ref{eq:Rates_full}) makes numerically evaluating Eq.~(\ref{eq:Rates_full}) computationally expensive. It is valuable to have a simplified expression for rates in this "low-field" regime, which for the proton corresponds to requiring 
\begin{equation}
 eB \ll (E_{Fp}-U_p)^2 - {m^*_p}^2 \ .
\label{eq:low-B-T0}
\end{equation}
At saturation density, this corresponds to $eB \ll 10^{18}\,\G$
When this criterion is met, the sum over Landau levels in Eq.~(\ref{eq:Rates_full}) can be replaced by the double integral:
\begin{equation}
    2eB\sum_{l\,l'}F^2_{l'\,l} \longrightarrow \int dk^2_{e\perp} dk^2_{p\perp} \dfrac{F^2_{l',l}(u)}{2eB}\, .
    \label{eq:replace_sum}
\end{equation}
As we will see in Sec.~\ref{sec:results}, these fields are still large enough to affect Urca rates. The rates
are dominated by thermally activated degrees of freedom around the Fermi surface. When the rate is unsuppressed by kinematic constraints, magnetic field effects will be important when the discretization of the energy into Landau levels is comparable to the thermal blurring of the Fermi surface. From Eq.~\eqref{eq:Ee} and \eqref{eq:Ep} we see that for the proton, which sets the limit because it is the
particle with the smallest Fermi energy, this corresponds to 
\begin{equation}
 eB \gtrsim (E_{Fp}-U_p)\, T \ . \label{eq:low-B}
\end{equation} 
At $T = 1$ MeV near saturation density this corresponds to $eB \gtrsim 10^{17}\,\G$. However, as the results presented in Sec.~\ref{sec:results} show, at densities where direct Urca processes would be kinematically forbidden at $B=T=0$ the field does not need to be this large in order to cause significant enhancement of the rate.

In the low-field regime we can write
$dk_{pz}\,dk_{p\perp}^2= d^3\mathbf{k}_p/\pi=2\,E_p^*\, k_p\, dE_p$ sin$\theta_p d\theta_p$ and $dk_{ez}\,dk_{e\perp}^2= d^3\mathbf{k}_e/\pi=2\,E_e\, dE_e\, k_e$ sin$\theta_e d\theta_e$. 
Thus the rate equation becomes
\begin{align}
    \Gamma_\text{nd}=&\frac{G^2(1+3g_A^2)}{4\pi^5} \int dk_n\, dk_p\, dk_e\, d\theta_p\, d\theta_e \,\Theta(E_{\nu}) \notag \\& E_{\nu}^2\, k_n\, k_p^2\, k_e^2\, f_n (1-f_p)(1-f_e) \text{sin}\theta_p\,\text{sin}\theta_e\, \label{eq:Rates_lowB}\\& \mathcal{F}(u)\times \Theta(k_n-\lvert k_p\text{cos}\theta_p+k_e\text{cos}\theta_e\rvert)\, \notag .
\end{align}
Here $\mathcal{F}(u)$ is the low-field limit of $F^2_{l',l}(u)/2eB$
which we now estimate.

\begin{figure}
  \centering
  \includegraphics[width=0.45\textwidth]{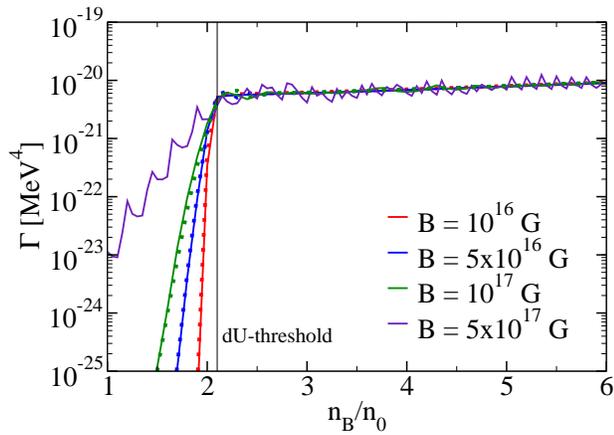}
   \caption{
  Neutron decay rate at low temperature $T=0.1$ MeV in the presence of finite magnetic field for the TMA EoS. The dots show the rate in the Fermi-surface approximation in the presence of low magnetic field using the formalism from \cite{Baiko:1998jq} and shown in Eqs.~(\ref{eq:Rates_Baiko}), (\ref{eq:Gamma0_Baiko}) and (\ref{eq:Rbqc_lowb}). The solid lines are rate calculations with the complete phase space integral  in the presence of low magnetic fields shown in Eq.~(\ref{eq:Rates_lowB}). The rate calculations match with the Fermi-surface approximation in the presence of a low magnetic field.}
   \label{fig:T0_Bneq0}
\end{figure}

From the approach described in Refs.~\cite{Kaminker1981, Riquelme:2005ac}, the range of $u$ can be divided into the following regimes: $u\ll u_1$, $u$ close to $u_1$, $u_1<u<u_2$,
$u$ close to $u_2$, $u\gg u_2$, where
\begin{equation}
\begin{array}{rl@{\quad}rl}
    u_{1,2}&=\dfrac{(k_{p\perp}\mp k_{e\perp})^2}{2eB}\ , & \Delta&=u_2-u_1 \ , \\[2ex]
   k_{p\perp} &= \sqrt{2eB l'} \ , & k_{e\perp} &= \sqrt{2eB l}\ .
\end{array}
\end{equation}
We now describe how $\mathcal{F}(u)$ is estimated in each of these regions.
When $u$ is far below $u_1$ or far above $u_2$ we have \cite{Kaminker1981, Riquelme:2005ac} 
\begin{align}
      u_1-u\gg (u_1^2/\Delta)^{1/3}:& \quad  \mathcal{F}(u) = \frac{e^{2u\sqrt{\lvert f(u)\rvert}-2\Phi}}{8\pi eB\, u\sqrt{\lvert f(u)\rvert}}\, , \label{eq:u_farfrom_u1}\\
   u-u_2\gg (u_2^2/\Delta)^{1/3}:& \quad    \mathcal{F}(u) = \frac{e^{-2u\sqrt{\lvert f(u)\rvert}-2\Phi}}{8\pi eB\, u\sqrt{\lvert f(u)\rvert}}\, ,\label{eq:u_farfrom_u2}
 \end{align}
where
 \begin{align}
     \Phi=& \frac{\sqrt{u_1u_2}}{2} \ln\!\left[\frac{\left(\sqrt{u_1\,\lvert u_2{-}u\rvert}+\sqrt{u_2\,\lvert u_1{-}u \rvert}\right)^2}{u\Delta}\right] \notag \\
     &+ \frac{u_1+u_2}{4}\ln\left[\frac{\left(\sqrt{\lvert u_2{-}u\rvert}-\sqrt{\lvert u_1{-}u \rvert}\right)^2}{\Delta}\right]\,  ,
 \end{align}
 \begin{equation}
     f(u)=\frac{1}{4u^2}(u-u_1)(u-u_2) \, .
 \end{equation}
 When $u$ is close to $u_1$, $\lvert u-u_1 \rvert\ll$ min$(u_1,\;\Delta)$,
  \begin{align}
     \mathcal{F} &= \left(\frac{4}{u_1\Delta}\right)^{\!\!1/3} \frac{(\text{Ai}(\xi_1))^2}{2eB}\, , \\ \xi_1&=(u_1-u)\left(\frac{\Delta}{4u_1^2}\right)^{1/3}\, ,
 \end{align}
 where Ai$(x)$ is the Airy function of the first kind. When $u$ is near $u_2$, $\lvert u-u_2\rvert\ll \Delta$, 
 \begin{align}
     \mathcal{F} &= \left(\frac{4}{u_2\Delta}\right)^{\!\!1/3} \frac{(\text{Ai}(\xi_2))^2}{2eB}\, \label{eq:u2_point}, \\
     \xi_2 &=(u-u_2)\left(\frac{\Delta}{4u_2^2}\right)^{\!\!1/3}\, .
 \end{align}
When $u$ is in the range $u_1 < u <u_2$, $\mathcal{F}$ is oscillatory. $\mathcal{F}$ averaged over oscillations is given by
 \begin{equation}
    \left< \mathcal{F} \right> = \frac{1}{2\pi eB\,\sqrt{u-u_1}\sqrt{u_2-u}}\, . 
    \label{eq:Fsq_allowed}
 \end{equation}
Using these expressions in \eqref{eq:Rates_lowB} we can obtain the neutron decay rate in low magnetic fields. 
The expression for electron capture is the same as Eqs.~(\ref{eq:Rates_full}), (\ref{eq:Rates_lowB}), but with phase space factors $f_n (1-f_p)(1-f_e)$ replaced by $(1-f_n) f_p\,f_e$ and $E_{\nu} = E_p+ E_e - E_n$.

\section{Consistency checks}
\label{sec:checks}

\begin{figure}
  \begin{subfigure}[t]{0.45\textwidth}
    \centering
    \includegraphics[width=\textwidth]{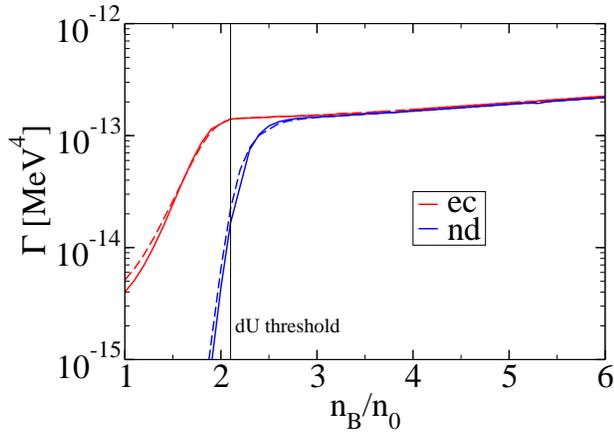}
    \caption{Rates at $T=3$ MeV.}
    \label{fig:T3_B0.06}
  \end{subfigure}\hfill
  \begin{subfigure}[t]{0.45\textwidth}
        \centering
    \includegraphics[width=\textwidth]{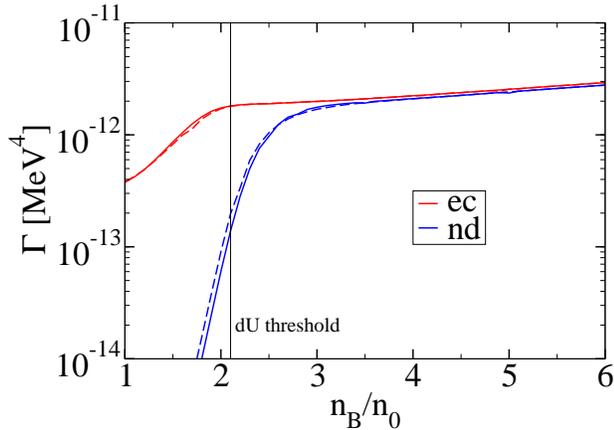}
    \caption{Rates at $T=5$ MeV.}
    \label{fig:T5_B0.06}

   \end{subfigure}
   \caption{Rates at finite temperatures in the presence of negligible finite magnetic fields for the TMA EoS. The dashed lines are rates calculated with the full phase-space integral as in \cite{Alford:2018lhf} in the absence of magnetic fields. The solid lines are rate calculations with the complete phase space integral in the presence of low magnetic fields following Eq.~(\ref{eq:Rates_lowB}).}
   \label{fig:Tneq0_B0}
\end{figure}

To check the consistency of our formalism with previously published results, we now present results for the first 3 cases in the enumeration given at the end of Sec.~\ref{sec:rates}.

\subsection{Case 1 : Low temperature and negligible magnetic field}

We compare our full phase space (no Fermi-surface approximation)
rate calculation at $B\rightarrow 0$ from Sec.~\ref{sec:low_field_rates} in the low temperature limit  with the  analytic expression for the rates in the Fermi-surface approximation at low temperatures in absence of magnetic field, taken from Eq.~(9) in Ref.~\cite{Alford:2018lhf},
\begin{equation}
    \Gamma_\text{nd}=\mathcal{A}_{dU}G^2 (1+3g_A^2)E_{F_n}^*\,E_{F_p}^*\,\mu_e\,T^5\Theta(k_{F_p}+k_{F_e}-k_{F_n}) \,,
    \label{eq:DU_FS_B0}
\end{equation} 
where $\mathcal{A}_{dU}=0.0170$ and $E_{F_N}^*=E_{F_N}- U_N$.
The rate calculations were found to agree within $2 \%$ for TMA EoS at temperature $T=0.1$ MeV and magnetic field $B=10^{12}\,\G$ over the density range $n_B=1-6\,n_0$.

\begin{figure*}
  \begin{subfigure}{0.33\textwidth}
    \centering
    \includegraphics[width=\textwidth]{Rates_T0.5_TMA_nd_scaled.eps}
    \caption{Rates at $T=0.5$ MeV}
    \label{fig:T0.5_TMA_nd}
  \end{subfigure}\hfill
  \begin{subfigure}{0.33\textwidth}
        \centering
    \includegraphics[width=\textwidth]{Rates_T1_TMA_nd_scaled.eps}
    \caption{Rates at $T=1$ MeV}
    \label{fig:T1_TMA_nd}

   \end{subfigure}\hfill
   \begin{subfigure}{0.33\textwidth}
        \centering
    \includegraphics[width=\textwidth]{Rates_T3_TMA_nd_scaled.eps}
    \caption{Rates at $T=3$ MeV}
    \label{fig:T3_TMA_nd}

   \end{subfigure}
   \caption{Direct Urca neutron decay rates for the TMA EoS. The vertical dashed line is the direct Urca threshold, given by $n_B=2.1n_0$ for this EoS. We observe an enhacement of the rate below the threshold, and only a minor dependence of the rate on the magnetic fields above the threshold. At the largest magnetic field plotted here, oscillations due to the De Haas–Van Alphen effect become visible. We plot the modified Urca rates at $B=0$ for comparison. At densities below threshold, we observe an enhancement of the rate above the modified Urca rate, highlighting the importance of our calculation in this regime. }
   \label{fig:ndrates_TMA}
\end{figure*}
\begin{figure*}
  \begin{subfigure}{0.33\textwidth}
    \centering
    \includegraphics[width=\textwidth]{Rates_T0.5_TMA_ec_scaled.eps}
    \caption{Rates at $T=0.5$ MeV}
    \label{fig:T0.5_TMA_ec}
  \end{subfigure}\hfill
  \begin{subfigure}{0.33\textwidth}
        \centering
    \includegraphics[width=\textwidth]{Rates_T1_TMA_ec_scaled.eps}
    \caption{Rates at $T=1$ MeV}
    \label{fig:T1_TMA_ec}

   \end{subfigure}\hfill
   \begin{subfigure}{0.33\textwidth}
        \centering
    \includegraphics[width=\textwidth]{Rates_T3_TMA_ec_scaled.eps}
    \caption{Rates at $T=3$ MeV}
    \label{fig:T3_TMA_ec}

   \end{subfigure}
   \caption{Direct Urca electron capture rates for the TMA EoS. The vertical dashed line is the direct Urca threshold, given by $n_B=2.1n_0$ for this EoS. }
   \label{fig:ecrates_TMA}
\end{figure*}

\subsection{ Case 2 : Low temperatures and non-negligible magnetic field }

We now compare our full phase space (no Fermi-surface approximation) rate calculation  in the low temperature limit for non-negligible magnetic field 
from Sec.~\ref{sec:rates}, \ref{sec:low_field_rates} with the
 analytic expression for the rates in the Fermi-surface approximation at low temperatures and low magnetic field, taken from~\cite{Baiko:1998jq}, summarised below.

Replacing particle momenta, $k_i$, with their Fermi momentum $k_{F_i}$ and particle energies, $E_i$ with Fermi energies $E_{F_i}$ in all the smooth functions in Eq.~(\ref{eq:Rates_lowB}) we have,
\begin{equation}
    \Gamma_\text{nd}=\Gamma_0 R_B^{qc}\, ,
    \label{eq:Rates_Baiko}
\end{equation}
where,
\begin{equation}
\begin{split}
    \Gamma_0 = &\frac{G^2 (1+3g^2_A)}{4\pi^5} E^*_{F_n}\, E^*_{F_p}\, \mu_e \int dE_{\nu} E^2_{\nu} dE_n \, dE_p \, dE_e\,\\& f_n (1-f_p) (1-f_e) \delta (E_n - E_p - E_e -E_{\nu}) \, ,
\end{split}
\end{equation}
which gives
\begin{equation}
    \Gamma_0 = \mathcal{A}_{dU}\times {G^2 (1+3g^2_A)} E^*_{F_n}\, E^*_{F_p}\, \mu_e \,T^5 \, ,
    \label{eq:Gamma0_Baiko}
\end{equation}
where $\mathcal{A}_{dU} = 0.0170$ as in Eq.~(\ref{eq:DU_FS_B0}) and,
\begin{equation}
\begin{split}
     R_B^{qc}=&\int_{-1}^1 d\,\text{cos}\theta_n\, d\,\text{cos}\theta_p\, d\,\text{cos}\theta_e \,k_{F_n}\,k_{F_p}\,k_{F_e}\frac{F^2_{l',l}(u)}{2eB} \\&\delta (k_{nz} -k_{pz} -k_{ez} )\, .
\end{split}
\end{equation}
By integrating over the polar angle of the neutron momentum we obtain
\begin{equation}
\begin{split}
    R_B^{qc}=&\int_{-1}^1 d\,\text{cos}\theta_p\, d\,\text{cos}\theta_e \,k_{F_p}\,k_{F_e}\frac{F^2_{l',l}(u)}{2eB} \\&\Theta (k_{F_n}-\lvert k_{F_p}\,\text{cos}\theta_p +k_{F_e}\,\text{cos}\theta_e\rvert ) \, .
     \label{eq:Rbqc_lowb}
\end{split}
\end{equation}
Following \cite{Baiko:1998jq} only the low-field asymptotic form of $\mathcal{F}$ given in Eq.~(\ref{eq:u2_point}) is used
in Eq.~(\ref{eq:Rbqc_lowb}) to calculate rates in this regime where the temperature effects are negligible and only particles with energies close to their Fermi energies participate in the reactions,
\begin{equation}
    \mathcal{F}=\frac{F^2_{l',l}}{2eB}= \frac{(k_p\text{sin}\theta_p k_e\text{sin}\theta_e)^{-1/3}}{(2eB)^{1/3}(k_p\text{sin}\theta_p +k_e\text{sin}\theta_e)^{2/3}}\, \text{Ai}^2(\xi)\, ,
    \label{eq:F_lowb}
\end{equation}
\begin{equation}
\begin{split}
    \xi=&\frac{1}{(2eB)^{2/3} (k_p\text{sin}\theta_p+k_e\text{sin}\theta_e)^{4/3}}\times\\& (k_n^2-k_p^2-k_e^2-2k_pk_e\text{cos}(\theta_p-\theta_e))\times\\&(k_p k_e \text{sin}\theta_p\text{sin}\theta_e)^{1/3}\,  ,
    \label{eq:xi}
\end{split}
\end{equation}
where we replace $k_i$ by $k_{F_i}$.

Comparing Eqs.~(\ref{eq:Rates_Baiko}) and (\ref{eq:Gamma0_Baiko}) with Eq.~(\ref{eq:DU_FS_B0}) and noting that $R^{qc}_B$ is non-zero even below the threshold where $k_{F_n}-k_{F_p}-k_{F_e} > 0$, we see that the direct Urca rates in the presence of magnetic fields are not suppressed strongly directly below the direct Urca threshold. Thus in the presence of magnetic fields, we see that direct Urca processes are allowed below the threshold density. In \cite{Baiko:1998jq} the sum over Landau levels is replaced by an integral, as in Eq.~(\ref{eq:replace_sum}) as the fields are assumed to be small enough such that the particles occupy a large number of Landau levels. This formalism is thus valid only for low magnetic fields, $B\leq  10^{17}\,\G$ such that $N_{F_p}=N_{F_e} > 10$, where $N_{F_i}=k^2_{F_i}/2eB$.

From Fig.~\ref{fig:T0_Bneq0} we see that our rate calculations agree well with the rates in the Fermi-surface approximation for low magnetic fields (dotted lines). In the presence of magnetic fields, Urca rates start becoming significant even below the direct Urca threshold. This effect becomes more pronounced for the larger fields.
\subsection{Case 3: High temperature and negligible magnetic field}

Here, we calculate rates for the full phase space at finite temperature for negligible magnetic fields using Eq.~(\ref{eq:Rates_lowB}) from Sec.\ref{sec:low_field_rates}. We set $T=3,\,5$ MeV as the rates deviate significantly from the Fermi-surface approximation at these temperatures as seen in \cite{Alford:2018lhf, Alford:2021ogv}. We choose $B=10^{12}\,\G$ as this magnetic field strength will have no effect on the Urca rates. As we consider finite temperatures here with negligible magnetic field, we can compare our rate calculations with those calculated in \cite{Alford:2018lhf} using the full phase space integral in the absence of an external magnetic field. Fig.~\ref{fig:Tneq0_B0} shows the rates calculated at temperatures $T=3,\, 5$ MeV and low magnetic field $B=10^{12}\,\G$. The dashed lines show the rates in the absence of magnetic field calculated as in \cite{Alford:2018lhf}. Our calculation of the rates at small magnetic fields agree with the rates calculated in the complete absence of magnetic fields.

\section{Results}
\label{sec:results}

We now show the results for case 4 in the enumeration given at the end of Sec.~\ref{sec:rates},  the regime where magnetic field effects are non-negligible and the Fermi-surface approximation is not valid. As mentioned in Sec~\ref{sec:formalism}, we perform this study for two different EoS: TMA and QMC-RMF3.
\begin{figure}
    \centering
    \includegraphics[width=0.45\textwidth]{Rates_ratio_vs_B_TMA.eps}
    \caption{The ratio of electron capture rates to neutron decay rates as a function of the magnetic field for the TMA EoS at the sub-threshold density $n_B = 1.8n_0$.}
    \label{fig:Rates_vs_B_TMA}
\end{figure}

\subsection{TMA}
We plot the direct Urca neutron decay (Fig.~\ref{fig:ndrates_TMA}) and electron capture  (Fig.~\ref{fig:ecrates_TMA}) rates for the TMA EoS at finite temperatures and magnetic fields.
This EoS has a direct Urca threshold at density $n_B=2.1n_0$.  We plot the rates at temperatures $T{=}0.5,\;1,\;3$ MeV and at magnetic fields $B{=}10^{16},\;5\times10^{16},\;10^{17},$ and $5\times10^{17}\,\G$.
We use the cold flavor equilibrium condition for matter where $\mu_n= \mu_p + \mu_e$.

For both neutron decay and electron capture we see that the magnetic field is only important below the direct Urca threshold density, where both nonzero temperature and nonzero magnetic field help to enhance the rate, blurring the dUrca threshold. At $T{=}0.5$ MeV a magnetic field $B{=}10^{16}\,\G$ is already more important than thermal blurring, and the enhancement increases as the field grows.
At $T{=}3$ MeV it takes a magnetic field of $5\times10^{17}
\,\G$ to produce further enhancement beyond the thermal blurring. For reference we have also plotted the standard 
(zero magnetic field) calculation of the modified Urca rate \cite{Alford:2018lhf, Alford:2021ogv}, so one can see that the enhancement will still be noticeable even when modified Urca processes are included.
At large magnetic fields, distinct De Haas–Van Alphen oscillations due to the increasing number of filled Landau levels become visible.

The equilibrium proton fraction  is determined by the balance between neutron decay and electron capture rates. In neutrino transparent matter at $T \gtrsim 1$ MeV it must be determined by explicitly calculating these rates and varying $x_p$ to find the value at which they balance \cite{Alford:2018lhf}.
In Fig.~\ref{fig:Rates_vs_B_TMA} we plot the ratio of electron capture rates to neutron decay rates as a function of increasing magnetic field at a density $n_B=1.8n_0$ and proton fraction $x_p=0.1$, which is below the direct-Urca threshold density for the TMA EoS. We observe that increasing the magnetic field reduces the difference between electron capture and neutron decay rates. This is because the neutron decay rates are more strongly enhanced by the magnetic field compared to the electron capture rates.
Thus magnetic fields are changing the chemical potentials at which the rates will balance each other and will therefore
affect the equilibrium proton fraction.

\begin{figure*}
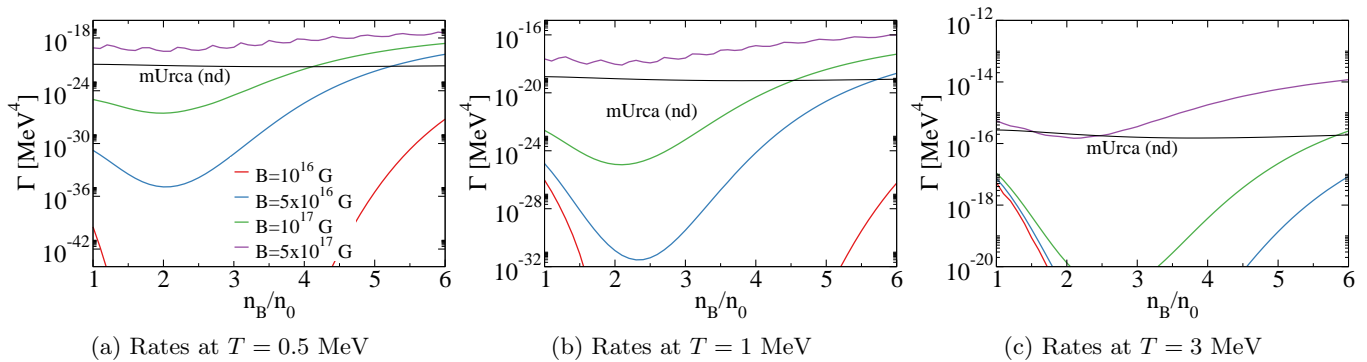
 
  \begin{subfigure}{0.33\textwidth}
    \centering
    \includegraphics[width=\textwidth]{Rates_T0.5_QMC-RMF3_nd_scaled.eps}
    \caption{Rates at $T=0.5$ MeV}
    \label{fig:T0.5_QMC-RMF3_nd}
  \end{subfigure}\hfill
  \begin{subfigure}{0.33\textwidth}
        \centering
    \includegraphics[width=\textwidth]{Rates_T1_QMC-RMF3_nd_scaled.eps}
    \caption{Rates at $T=1$ MeV}
    \label{fig:T1_QMC-RMF3_nd}

   \end{subfigure}\hfill
   \begin{subfigure}{0.33\textwidth}
        \centering
    \includegraphics[width=\textwidth]{Rates_T3_QMC-RMF3_nd_scaled.eps}
    \caption{Rates at $T=3$ MeV}
    \label{fig:T3_QMC-RMF3_nd}

   \end{subfigure}
   \caption{Direct Urca neutron decay rates for the QMC-RMF3 EoS.}
   \label{fig:ndrates_QMC-RMF3}
\end{figure*}

\begin{figure*}
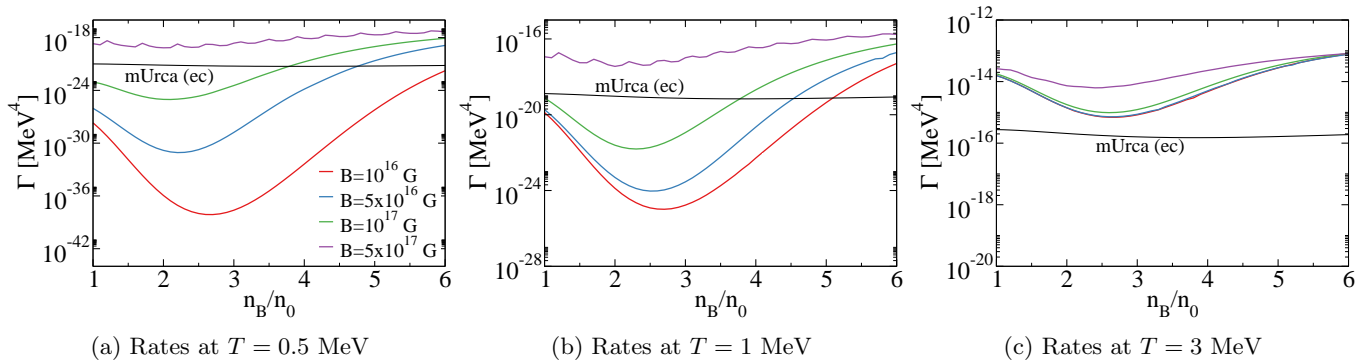

  \begin{subfigure}{0.33\textwidth}
    \centering
    \includegraphics[width=\textwidth]{Rates_T0.5_QMC-RMF3_ec_scaled.eps}
    \caption{Rates at $T=0.5$ MeV}
    \label{fig:T0.5_QMC-RMF3_ec}
  \end{subfigure}\hfill
  \begin{subfigure}{0.33\textwidth}
        \centering
    \includegraphics[width=\textwidth]{Rates_T1_QMC-RMF3_ec_scaled.eps}
    \caption{Rates at $T=1$ MeV}
    \label{fig:T1_QMC-RMF3_ec}

   \end{subfigure}\hfill
   \begin{subfigure}{0.33\textwidth}
        \centering
    \includegraphics[width=\textwidth]{Rates_T3_QMC-RMF3_ec_scaled.eps}
    \caption{Rates at $T=3$ MeV}
    \label{fig:T3_QMC-RMF3_ec}

   \end{subfigure}
   \caption{Direct Urca electron capture rates for the QMC-RMF3 EoS. }
   \label{fig:ecrates_QMC-RMF3}
\end{figure*}

\subsection{QMC-RMF3} 
We also studied Urca rates for the QMC-RMF3 EoS, which does not have a direct Urca threshold in the density range of interest. We again impose the cold flavor equilibrium condition on matter, $\mu_n=\mu_p+\mu_e$. The results are shown in Figs.~(\ref{fig:ndrates_QMC-RMF3}) and (\ref{fig:ecrates_QMC-RMF3}). 
We see that the Urca rates are always enhanced by the magnetic field as well as the temperature, with even greater sensitivity to the magnetic field than was seen for the TMA EoS.
A magnetic field $B{=}5{\times}10^{17}\,\G$ is easily enough to push the dUrca rate above the (zero field) mUrca rate, and at some densities and temperatures. e.g. for electron capture at $T=1$ MeV, even much smaller fields of order $10^{16}\,\G$ can accomplish this.
This may have important consequences for
physical processes where Urca rates are essential, such as neutron star cooling \cite{Page:2004fy,Page:2009fu}, transport in neutron stars and neutron star mergers \cite{Yakovlev:2000jp, Gavassino:2020kwo, Radice:2021jtw, Foucart:2022bth, Gavassino:2023xkt} and  for the thermal states of compact stars in
low-mass X-ray binaries \cite{Yakovlev:2004iq,Fortin:2017rxq,Shternin:2018dcn}.

\section{Conclusions}
\label{sec:discussions}

In this work, we investigated the effect of magnetic fields on the direct Urca rates in dense nuclear matter at temperatures relevant to neutron star mergers and proto-neutron stars. 
We performed calculations for matter described by two EoSs,  TMA (which has a direct Urca threshold) andd QMC-RMF3 (which does not).

We find that the effect of the magnetic fields is significant at relatively low temperatures and can be washed out by large temperatures. For temperatures around $T{=}1$ to $3$ MeV the neutron decay rate below the direct Urca threshold is enhanced enough to be comparable to modified Urca for fields of order $10^{16}$-$10^{17}\,\G$. The electron capture rate
is more strongly affected by nonzero temperatures, see e.g.~Ref.~\cite{Alford:2021ogv},
so for that process the effects of magnetic fields of order  $10^{16}$-$10^{17}\,\G$ are less obvious at $T{=}3$ MeV, but become noticeable at $T=0.5$ MeV.

We see that with increasing magnetic field the difference between direct Urca neutron decay and electron capture rates is reduced (Fig.~\ref{fig:Rates_vs_B_TMA}). This can reduce the isospin chemical potential required for the correct criterion of flavor equilibrium at finite temperatures shown in \cite{Alford:2018lhf,Alford:2021ogv, Alford:2023gxq}. 

We performed the study for two different EoSs and found that the magnetic field counteracts the suppression of
Urca rates at densities where Urca is kinematically forbidden at $T=B=0$, e.g.~at densities below the direct Urca threshold.
The magnetic field effects were most significant for the QMC-RMF3 EoS, for which direct Urca is forbidden at all densities relevant to neutron stars, and for the TMA EoS at densities below its dUrca threshold.

The scheme applied in this work for the calculation of dUrca rates is valid for moderate magnetic fields, $B<10^{18}\,\G$, for which a large number of Landau levels are occupied (see Sec.~\ref{sec:low_field_rates}). This is not applicable to the case when the magnetic fields are very large $B> 10^{18}\,\G$, in which case only the lowest Landau level may be occupied. However, it is expected that such large magnetic fields will not arise in neutron star mergers and proto-neutron stars, the upper limit being around  $B\sim 10^{17}\,\G$~\cite{Lai1991, Kiuchi:2017zzg,Palenzuela:2021gdo,Kiuchi:2023obe}. 

The results of this study may have important implications for astrophysical scenarios where magnetic fields in the $10^{16}$-$10^{17}\,\G$ range are present and Urca processes play an important role, e.g.~in flavor equilibration of the proton fraction, neutrino opacity, bulk viscosity, neutron star cooling, etc.

The formalism used in this work could be combined with the recently developed nucleon width approximation (NWA)  \cite{Alford:2024xfb} to calculate magnetic field and finite temperature corrections to in-medium effects (corresponding to modified Urca contributions to the rate).
Furthermore, it would be interesting to investigate the effect of magnetic fields on the Urca rates in neutron star mergers, as has been done in the zero magnetic field case in Refs.~\cite{Most:2022yhe, Espino:2023dei}. Our calculations also allow us to study the cooling behavior of (proto-) neutron stars, where a ``softening" of the direct Urca threshold could lead to faster cooling.

\begin{acknowledgments}

P.T. thanks Suprovo Ghosh and Nilaksha Barman for helpful discussions on the topic. D.C. acknowledges Lami Suleiman and Arus Harutyunyan for insightful discussions. D.C. is thankful for the warm hospitality of Prof. Mark Alford, Alexander Haber and their research group at the Washington University in St. Louis. P.T. and D.C. acknowledge the usage of IUCAA HPC computing facility. MGA and AH are partly supported by the U.S. Department of Energy, Office of Science, Office of Nuclear Physics, under Award No.~\#DE-FG02-05ER41375.
\end{acknowledgments}
\clearpage  
\bibliography{apssamp}

\end{document}